\documentclass[twocolumn,showpacs,preprintnumbers,amsmath,amssymb]{revtex4}
\usepackage{graphicx}
\usepackage{dcolumn}
\usepackage{bm}
\usepackage{latexsym}
\usepackage{epsfig}
\begin{document}
\preprint{APS/123-QED}
\title{Three-body break-up in deuteron-deuteron scattering at 65~MeV/nucleon}
\author{\firstname{A.}~\surname{Ramazani-Moghaddam-Arani$^{1,2}$}}
\hspace{-5cm}\email{ramazani@kvi.nl}
\author{\firstname{H.R.}~\surname{Amir-Ahmadi$^{2}$}}
\author{\firstname{A.D.}~\surname{Bacher$^{3}$}}
\author{\firstname{C.D.}~\surname{Bailey$^{3}$}}
\author{\firstname{A.}~\surname{Biegun$^{2}$}}
\author{\firstname{M.}~\surname{Eslami-Kalantari$^{2,4}$}}
\author{\firstname{I.}~\surname{Ga\v{s}pari\'c$^{5}$}}
\author{\firstname{L.}~\surname{Joulaeizadeh$^{2}$}} 
\author{\firstname{N.}~\surname{Kalantar-Nayestanaki$^{2}$}}
\author{\firstname{St.}~\surname{Kistryn$^{6}$}}
\author{\firstname{A.}~\surname{Kozela$^{7}$}}
\author{\firstname{H.}~\surname{Mardanpour$^{2}$}}
\author{\firstname{J.G.}~\surname{Messchendorp$^{2}$}}\email{messchendorp@kvi.nl}
\author{\firstname{A.M.}~\surname{Micherdzinska$^{8}$}}
\author{\firstname{H.}~\surname{Moeini$^{2}$}}
\author{\firstname{S.V.}~\surname{Shende$^{2}$}}
\author{\firstname{E.}~\surname{Stephan$^{9}$}}
\author{\firstname{E.J.}~\surname{Stephenson$^{3}$}}
\author{\firstname{R.}~\surname{Sworst$^{6}$}}
\affiliation{%
$^1$ Department of Physics, Faculty of Science, University of Kashan, Kashan, Iran}
\affiliation{%
$^2$ KVI, University of Groningen, Groningen, The Netherlands}
\affiliation{%
$^3$ Center for Exploration of Energy and Matter, Indiana University, Bloomington, Indiana, USA}%
\affiliation{%
$^4$ Department of Physics, Faculty of Science, Yazd University, Yazd, Iran}
\affiliation{%
$^5$ Rudjer Bo\v{s}kovi\'c Institute, Zagreb, Croatia}
\affiliation{%
$^6$ Institute of Physics, Jagiellonian University, Cracow, Poland}
\affiliation{%
$^7$ Henryk Niewodnicza\'nski, Institute of Nuclear Physics, Cracow, Poland}
\affiliation{%
$^8$ University of Winnipeg, Winnipeg, Canada}%
\affiliation{%
$^9$ Institute of Physics, University of Silesia, Katowice, Poland}%
 \date{\today}
\begin{abstract}
In an experiment with a 65 MeV/nucleon polarized deuteron beam on a
liquid-deuterium target at KVI, several multi-body final states in
deuteron-deuteron scattering were identified. For these measurements,
a unique and advanced detection system, called BINA, was utilized.  We
demonstrate the feasibility of measuring vector and tensor
polarization observables of the deuteron break-up reaction leading to
a three-body final-state.  The polarization observables were
determined with high precision in a nearly background-free
experiment. The analysis procedure and some  results are
presented.
\end{abstract}
 \maketitle
\section{Introduction}
The physics phenomena of nuclei are largely understood by considering
the interaction between their building blocks, the nucleons. In 1935,
Yukawa described the nucleon-nucleon (NN) force by the exchange of
massive mesons~\cite{Yukawa} in analogy to the electromagnetic
interaction which can be represented by the exchange of a massless
photon. Several phenomenological nucleon-nucleon potentials have been
derived based on Yukawa's theory and are able to reproduce data points
in neutron-proton and proton-proton scattering with extremely high
precision. These so-called high-quality NN potentials are used in
Faddeev equations~\cite{RamazaniA_Ref2,RamazaniA_Ref3} to give an
exact solution of the scattering problem for the three-nucleon
system. Already, for the simplest three-nucleon system, the triton, an
exact solution of the three-nucleon Faddeev equations employing
two-nucleon forces (2NFs) underestimates the experimental binding
energy~\cite{RamazaniA_Ref4}, showing that 2NFs are not sufficient to
describe the three-nucleon system accurately. The existence of an
additional force, the three-nucleon (3N) interaction, was predicted a long time ago by
Primakov~\cite{RamazaniA_Ref5} and confirmed by a comparison between
precision data and state-of-the-art calculations~\cite{Wit98}.

Many high-precision measurements of nucleon-deuteron scattering
processes at intermediate energies were carried out in the past
decades with the aim to study 3NF effects and to compare the
experimental observations with the predictions from rigorous Faddeev
calculations. In general, adding 3NF effects to the modern NN
potentials gives a better agreement between the cross section data for
the proton-deuteron scattering processes and the corresponding
calculations~{\cite{RamazaniA_Ref9,RamazaniA_Ref10,RamazaniA_Ref11,RamazaniA_Ref12,RamazaniA_Ref13,RamazaniA_Ref14,RamazaniA_Ref15,RamazaniA_Ref16,RamazaniA_Ref17,Ahmad07,Ahmad08}},
whereas a similar comparison for the spin observables yields various
discrepancies~{\cite{RamazaniA_Ref11,RamazaniA_Ref13,RamazaniA_Ref15,Wit93,Ed99,Mesh00,RamazaniA_Ref18,Cadman01,Kuros02,Seki02,Hatanaka02,Meyer04,RamazaniA_Ref19,Mehman05,Przew06,Kist06,Bieg06,RamazaniA_Ref21,Stephan07,RamazaniA_Ref20,RamazaniA_Ref22,Mardan10}}.
The overall conclusion is that the spin-dependent parts of the 3NFs
are poorly understood and that more studies in this field are needed.

The 3NF effects are in general small in the three-nucleon system. A
complementary approach is to examine heavier systems for which the 3NF
effects are significantly enhanced in magnitude.  Naively, one might
expect that the 3NF effects increase by the argument that the number
of three-nucleon combinations with respect to two-nucleon combinations
gets larger with increasing number of nucleons. We, however, note that
the saturation of 3NF effects sets in very quickly for large nuclei as
well.  This simple counting rule is supported by a comparison between
predictions and data for the binding energies of light
nuclei~\cite{Pieper01}. The predictions of a Green's function
Monte-Carlo calculation based on the Argonne V18~\cite{AV18} NN
interaction (AV18) and the Illinois-2 (IL2) 3NF~\cite{Pudliner95} are
compared to experimental data. While a calculation which only includes
the AV18 NN potential deviates significantly from the experimental
results, the predictions of calculations which include as well a 3NF
come much closer to the data, especially for the first few light
nuclei. Note that the effect of the 3NF on the binding energy for the
triton is 0.5-0.8~MeV, whereas the effect increases significantly for
the four-nucleon system, $^{4}$He, to $\sim$4~MeV.  For this, it was
proposed to study the four-nucleon system at intermediate energies
since the experimental database for this is presently poor in
comparison with that of three-nucleon
systems~{\cite{RamazaniA_Ref27,RamazaniA_Ref28,RamazaniA_Ref29,Mich2007}}.
Most of the available data have been measured at very low energies, in
particular below the three-body break-up threshold of 2.2~MeV. Also,
theoretical developments are evolving rapidly at low
energies~\cite{RamazaniA_Ref23,RamazaniA_Ref24,RamazaniA_Ref25,RamazaniA_Ref26},
but lag behind at higher energies. This situation calls for extensive
four-nucleon studies at intermediate energies.

This paper addresses the feasibility of measuring polarization
observables in the three-body break-up channel in deuteron-deuteron
scattering, $\vec d+d\rightarrow d + p + n$, for an incident beam
energy of 65~MeV/nucleon.  This work is part of a more extensive
experimental program which was carried out at KVI and that aims to
provide precision data in various four-nucleon scattering processes at
intermediate energies. With these data, we will significantly enrich
the four-nucleon scattering database at intermediate energies and,
thereby, provide a basis to test future calculations with the
long-term aim to understand the details of 3NF effects.

\begin{figure}[!htp]
\centering
\includegraphics[angle=0,width=.6\textwidth]{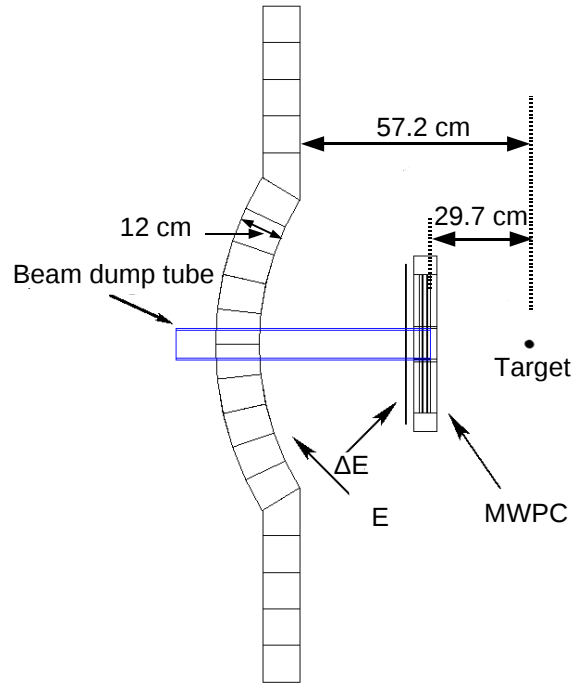}
\caption{(Color online) A side view of the forward part of BINA  which consists of a Multi-Wire Proportional Chamber
(MWPC), and an array of thin plastic scintillators $\Delta$E, followed
by a wall of thick segmented scintillators (E). The backward ball is not shown in this picture,
since it was not used in the analysis presented in this paper.}
\label{bina1} 
\end{figure}

\section{Experiment and experimental setup}
Deuteron-deuteron scattering below the pion-production threshold leads to 5 
possible final states with a pure hadronic signature, namely:\\ 
\noindent
\begin{enumerate}
\item Elastic channel: $ \vec d + d \longrightarrow d + d$ ;
\item Neutron-transfer channel:  $ \vec d + d \longrightarrow p + t$ ;
\item Proton-transfer channel:   $ \vec d + d \longrightarrow n +  ^{3}\!\rm He$ ;
\item Three-body final-state break-up: $ \vec d + d \longrightarrow p +\\ n + d$ ;
\item Four-body final-state break-up: $ \vec d + d \longrightarrow p + n + p + n$. 
\end{enumerate}
The study and identification of these final states require an
experimental setup with specific features, namely, a large phase space
coverage, a good energy and angular resolution, and the ability for
particle identification. The experiment presented in this paper was
carried out at KVI using the Big Instrument for Nuclear-polarization
Analysis, BINA, which meets all these
requirements~\cite{RamazaniA_Ref20,RamazaniA_Ref30}. This work is devoted to the
three-body final state.

BINA is a setup with a nearly $4\pi$ geometrical acceptance and has
been used in various few-nucleon scattering experiments to measure the
scattering angles and energies of protons and deuterons with the
possibility for particle identification. The detector is composed of a
forward part (forward wall) and a backward part. Figure~\ref{bina1}
shows a side view of the forward part of BINA which consists of a
Multi-Wire Proportional Chamber (MWPC), and an array of thin plastic
scintillators ($\Delta$E) with a thickness of 1~mm, followed by a wall
of thick segmented scintillators (E) with a thickness of 12~cm
each. All scintillators were read out by photo-multiplier tubes (PMTs)
on both ends.  The signals from these PMTs are integrated using a QDC
to measure their energy loss in the scintillator. The discriminated
signals from the Constant Fraction Discriminators (CFDs) are fed into
Time-to-digital Converters (TDCs). The time difference between the
signals from both ends of the scintillators allows to determine the
position of the incoming particle. In addition, information about the
time-of-flight (TOF) of the particles can be extracted.  The thick
scintillators were mounted in a cylindrical shape, thereby, facing the
target. The thickness of these scintillators is sufficient to stop all
the protons and deuterons originating from the processes described in
this work.  The forward wall has a complete azimuthal coverage for
scattering angles from 10$^\circ$ to 32$^\circ$, which is advantageous
for the determination of the beam polarization and spin observables.
With the wire spacing of 2 mm and for a distance of 29.7~cm between
MWPC and target, we obtain an overall accuracy of 0.4$^{\circ}$ for
the polar angle and between 0.6$^{\circ}$ and 2.0$^{\circ}$ for the
azimuthal angle.  The detection efficiency of the MWPC was obtained by
using an unbiased and nearly background-free data sample of
elastically-scattered deuterons and was found to be typically 98\%
with an absolute uncertainty of 1\%.

\begin{figure}[!h]
\centering 
\includegraphics[angle=0,width=.52\textwidth]{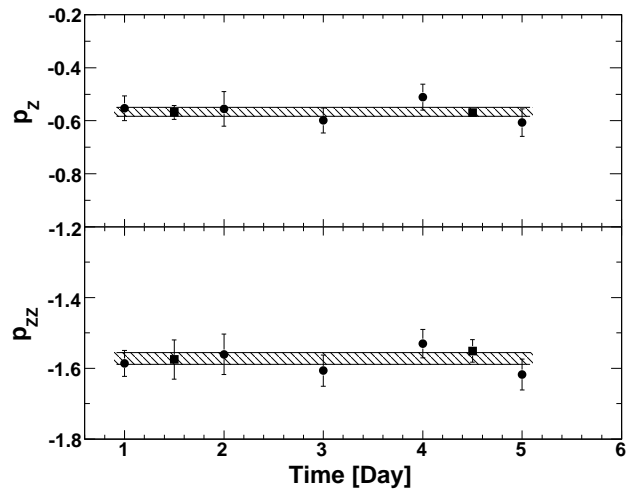} 
\caption{The results for the vector (top panel) and tensor (bottom
panel) polarizations of the deuteron beam. The filled circles and squares
are the measured values by BINA and LSP, respectively. The shaded bands
represent the result of a constant-value fit through all the data points.}
\label{resultpol}
\end{figure}

A polarized beam of deuterons originating from the polarized ion
source (POLIS)~\cite{Polis} was accelerated by AGOR
(Acc\'{e}l\'{e}rateur Groningen ORsay) to a kinetic energy of
65~MeV/nucleon and bombarded on a liquid-deuterium
target~\cite{RamazaniA_Ref31}.  The polarization of the deuteron beam
was measured in the low-energy beam line with a Lamb-Shift Polarimeter
(LSP)~\cite{Kremers1} as well as in the high-energy beam line with
BINA~\cite{RamazaniA_Ref30}.  For the polarization measurements using
the LSP, the low-energy beam was decelerated and focused on to the LSP
detection system. The number of deuterons in the spin-up state,
N$_{+}$, spin-zero state, N$_{0}$, and spin-down state, N$_{-}$, were
counted to measure directly the vector and tensor polarizations of the
beam.  In addition, the polarization of the beam of deuterons was
obtained at the high-energy beam line employing BINA via
$\phi$-asymmetry measurements of the $^{1}{\rm H}(\vec d,dp)$ reaction
based on its well-known analyzing power. Note that the same detector
was used as well for the measurement of the polarization observables
of the deuteron-deuteron break-up reaction.  With the self-calibrating
BINA setup, regular asymmetry measurements of the elastic
deuteron-proton scattering process were performed during the actual
experiment by switching rapidly the liquid target to a solid CH$_2$
target.  The polarization of the beam of deuterons was studied as a
function of time to check the stability of polarization during the
measurement. Also, the polarization values by BINA were compared with
those measured with the LSP. The results for the vector (top panel)
and tensor (bottom panel) polarizations of the deuteron beam from BINA
and LSP are shown in Fig.~\ref{resultpol} as a function of time. Each
filled circle represents the average value of the measured
polarizations deduced from different scattering angles with BINA in a
range from 55$^\circ$ to 140$^\circ$ in the center-of-mass frame. The
corresponding analyzing powers were taken from an angle-dependent fit
of literature
values~\cite{RamazaniA_Ref17,Stephan07,Wit93}. Polarization values
obtained for different scattering angles were found to be in good
agreement within uncertainties. The values of the polarization
obtained with LSP are depicted as filled squares.  The shaded bands
represent the results of a constant-value fit through the data
including the results obtained with BINA and LSP. The width of the
band corresponds to a 2$\sigma$ error of the fit. The polarization
values were found to be $p_{Z} = -0.57 \pm 0.03$ and $p_{ZZ} = -1.57
\pm 0.03$.  The total uncertainties of the polarizations was estimated
by adding quadratically the statistical error and the systematic
error. The systematic uncertainty stems from the uncertainty in the
analyzing powers, and was found to be 6\% and 5\% for the vector and
tensor polarizations, respectively.  A comparison between the results
demonstrates that the beam polarization was stable during the
experiments and that there is a good agreement between the measured
values obtained at the low-energy beam line with the LSP and those
measured at the high-energy beam line with BINA. The beam current was
monitored during the experiments via a Faraday cup at the end of the
beam line. The Faraday cup is made of a copper block containing a
heavy alloy metal as the actual beam stopper. The current meter was
connected to the Faraday cup with a short cable to avoid the voltage
drop and pickup effect. The current meter was calibrated using a
precision current source with an uncertainty of 2\%. The beam current
was typically 4 pA during the experiment. An offset of 0.24$\pm$0.07
pA in the beam current was observed in the $\vec dd$ experiment
through a comparison of $T_{20}$ in elastic scattering, measured with
BINA and the one measured with another experimental setup
(BBS)~\cite{BBS_ref}.

\section{Analysis method and event selection}
The elastic channel, neutron transfer channel, and break-up channels
leading to three- and four-body final states were uniquely identified
using the information on the energies of the outgoing particles, their
scattering angles, and their TOF. In this work, the deuteron-deuteron
break-up into the three-body final-state (four-body final-state) is
further referred to as the three-body break-up (four-body
break-up). This is the first time that spin observables in the
three-body break-up reaction in $\vec d+ d$ scattering process at
intermediate energies have been measured in a background-free
experiment. The analysis procedure and a part of the results of the
three-body break-up channel are presented in this paper.

The kinematics of the three-body break-up reaction are determined
unambiguously by exploiting the scattering angles of the proton and
the deuteron ($\theta_{d}, \theta_{p}, \phi_{12}=|\phi_{d}-\phi_{p}|$)
and the relation between their energies presented by the kinematical
curve which is referred to as the $S$-curve. The angles $\theta_{p}$
and $\theta_{d}$ are the polar angles of the proton and the deuteron,
respectively, and $\phi_{12}$ is the difference between their
azimuthal angles. The energies of the proton, $E_p$, and deuteron,
$E_d$, were transferred into two new variables, $D$ and $S$. The
variable $S$ is the arc-length along the $S$-curve with the starting
point chosen arbitrarily at the point where E$_d$ is minimum and $D$
is the distance of the ($E_p$, $E_d$) point from the kinematical
curve.
\begin{figure}[!h]
\centering
\includegraphics[width=1\columnwidth,angle=0]{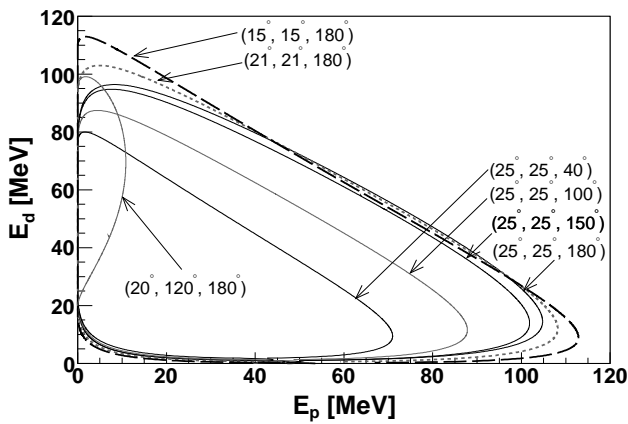}
\caption{The energy correlation between protons and deuterons in
  coincidence for the three-body break-up reaction in $\vec {d} + d$
  scattering is shown as $S$-curves for several kinematical
  configurations.  The kinematics are defined by ($\theta_{d},
  \theta_{p}, \phi_{12}$), the polar scattering angles of the deuteron
  and proton, respectively, and their relative azimuthal angle.}
\label{kinema}       
\end{figure}

The $S$-curves for several kinematical configurations are shown in
Fig.~\ref{kinema}. Each $S$-curve is labeled by three numbers. For
example, the label ($20^\circ, 120^\circ, 180^\circ$) shows a
kinematical relation of energies of a deuteron that scatters to
$20^{\circ}$ and a proton that scatters to $120^{\circ}$, and the
azimuthal opening angle, $\phi_{12}$, between them is equal to
$180^{\circ}$.

For the analysis of the three-body break-up data we measured the
kinetic energies and the polar and azimuthal angles of the two
coincident charged particles.  We performed the analysis of the
three-body break-up reaction for that part of the phase-space in which
the deuteron and the proton were detected in coincidence in the
forward part of BINA.  

For some parts of the $\Delta$E detector, the yield 
of scintillation light reaching the photomultipliers was not sufficient 
to obtain a good particle identification.
We, therefore, decided to make use of the time-of-flight to 
perform a particle identification instead. We note that the procedure 
was checked by a successful comparison with 
an analysis using those $\Delta$E detectors which were operating 
according to specifications. For the
particle identification, we measured the TOF of each registered
particle and compared its result with the expectation from the
kinematics of the three-body break-up reaction.  The TOF was
determined by discriminating the signals from the two PMTs from each
scintillator. The output of the constant-fraction discriminators were
fed into time-to-digital converters (TDCs) which were used in a
common-stop mode. The common stop signal was derived from the radio
frequency of the cyclotron. For the analysis of the break-up data, the
TDC output corresponding to the left- and right-hand side PMTs, $\rm
{TOF_L}$ and $\rm {TOF_R}$, were added together for each event.  The
sum of $\rm {TOF_L}$ and $\rm {TOF_R}$ is independent of hit position
along the scintillator slab.  We call this sum TOF$_i$ with $i$
referring to the different particles that hit the forward wall. For
particle identification, we compared these results with the expected
TOF that was calculated from the energy of a particle, calculating a
path length from a scattering angle and assuming a certain particle
type.  More precisely, the difference between the measured TOF of
particles 1 and 2 from the information of the TDCs, $(\rm{TOF}_1 -
\rm{TOF}_2)_{\rm{TDC}}$, and that extracted from the energies and the
scattering angles, $(\rm{TOF}_1 - \rm{TOF}_2)_{E}$, has been used to
define the variable $\Delta$TOF.  We note that $\Delta$TOF is a
difference of differences, and that it compares data with a kinematics
calculations, both of which yield a difference.  The identification of
the break-up channels proceeds by analyzing $\Delta$TOF.

\begin{figure}[!h]
\centering
\includegraphics[width=1\columnwidth,angle=0]{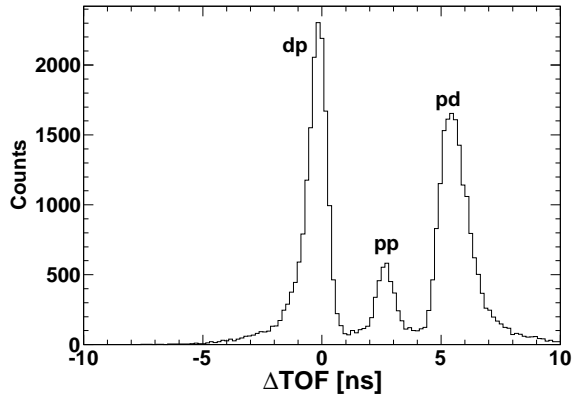}
\caption{The measured $\Delta$TOF for two particles that are detected
  in coincidence in the forward wall. The scattering angle of both
  particles is fixed to be $25^{\circ} \pm 2^{\circ}$ and the
  difference between the azimuthal angles of the two particles is
  $180^{\circ} \pm 5^{\circ}$.}
\label{TOF_TDC_E}
\end{figure}

 Figure~\ref{TOF_TDC_E} shows the value of $\Delta$TOF for two
 particles that are detected in coincidence in the forward wall. The
 scattering angle of both particles is fixed to be $25^{\circ} \pm
 2^{\circ}$ and the difference between the azimuthal angles of the two
 particles is $180^{\circ} \pm 5^{\circ}$. We note that $\Delta$TOF
 does not depend upon $S$ and that the detected particle in each hit
 can be a proton or a deuteron. Three clear peaks can be recognized
 corresponding to proton-deuteron, proton-proton, and deuteron-proton
 coincidences.  The identification of the peaks were confirmed using
 the $\Delta$E responses.  For this spectrum, we assumed in the
 calculation of $(\rm{TOF}_1 - \rm{TOF}_2)_{E}$ that the first
 particle is a deuteron and the second one a proton. If our assumption
 is correct, events which correspond to a deuteron-proton combination
 will give a peak around zero in the $\Delta$TOF spectrum. Note that a
 clear signal (peak on the left-hand side of Fig.~\ref{TOF_TDC_E}) of
 the three-body break-up events can be observed corresponding to a
 final-state consisting of a deuteron-proton combination. The peak on
 the most right-hand side corresponds to three-body break-up events,
 but with a proton-deuteron combination in the final state. The peak
 in the middle of the spectrum has been identified as two protons that
 stem from the four-body break-up reaction.

\begin{figure*}[!t]
\centering
\includegraphics[width=1.5\columnwidth,angle=0]{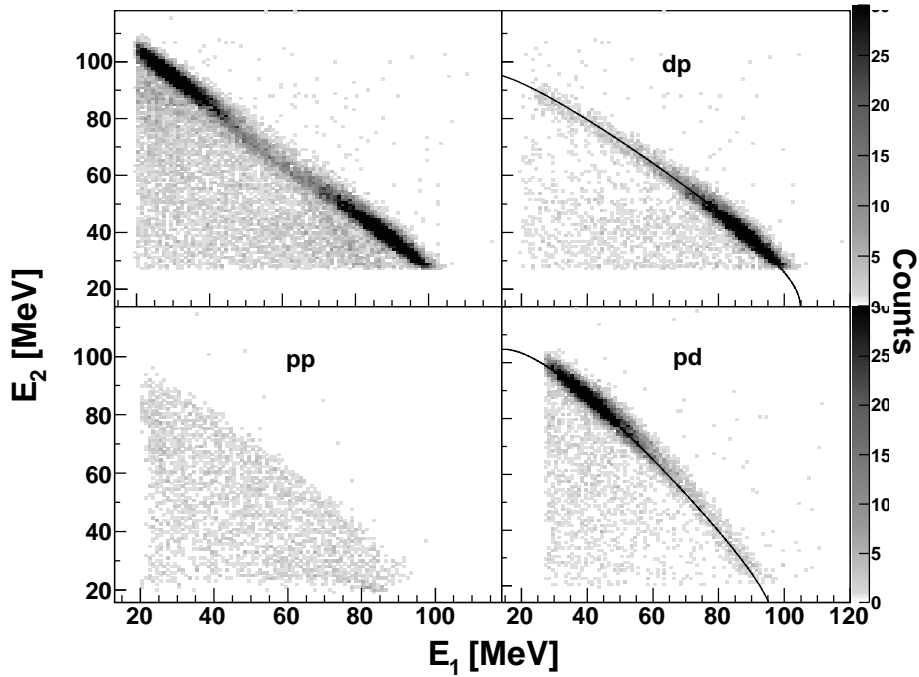}
\caption{The correlation between the calibrated energy of the two 
outgoing particles (proton or deuteron) in the deuteron-deuteron
break-up process is shown for the coplanar configuration,
($\theta_{1}, \theta_{2},
\phi_{12}) = (25^{\circ}, 25^{\circ}, 180^{\circ}$) before PID (top left
panel) and after PID (other panels). The bottom-panel on the left-hand
side depicts events from the four-body break-up reaction and the two
panels on the right-hand side correspond to the three-body break-up
channels.  The solid curves represent the relativistic $S$-curve that
are calculated from the energy and momentum conservation for the selected
configuration.}
\label{ChanID_tdc}
\end{figure*}
The top-left panel of Fig.~\ref{ChanID_tdc} represents the correlation
between the energy of two particles that are detected in coincidence
in the forward wall. The scattering angle of both particles is fixed
to be $25^{\circ} \pm 2^{\circ}$ and the difference between the
azimuthal angles of the two particles is $180^{\circ} \pm 5^{\circ}$.
The energies were obtained by a charge-integration of the signals from
the PMTs of the forward scintillators.  This spectrum contains events
from two different reactions, namely three- and four-body break-up
reactions.  The proton-deuteron or deuteron-proton coincidences from
the three-body break-up reaction can be separated by choosing events
corresponding to the peak of interest as shown in
Fig~\ref{TOF_TDC_E}. The results are shown in the top-right and
bottom-right panels. The bottom-left panel contains events which
correspond to the peak in the middle of Fig~\ref{TOF_TDC_E}. They
originate from the four-body break-up channel. Figures~\ref{TOF_TDC_E}
and \ref{ChanID_tdc} demonstrate the particle identification
capabilities of the detection system.

\begin{figure}[!t]
\centering
\includegraphics[width=1\columnwidth,angle=0]{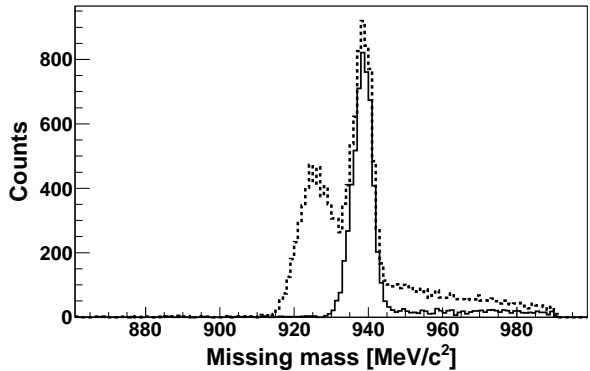}
\caption{The calculated mass of the undetected particle in the
  three-body break-up reaction in deuteron-deuteron scattering. The
  energy, polar and azimuthal angles of the detected protons are used
  in the calculation.  The dashed (solid) curve represents value of
  the calculated mass of the missing particle before (after) applying
  the identification cut in Fig.~\ref{TOF_TDC_E}. The scattering angle
  of both detected particles is fixed to be $25^{\circ} \pm 2^{\circ}$
  and the difference between the azimuthal angles of the two detected
  particles is $180^{\circ} \pm 5^{\circ}$. }
\label{missmass}       
\end{figure}

The identification of the proton and deuteron enables us to measure
the missing particle in the three-body break-up reaction in
deuteron-deuteron scattering.  The missing particle in this reaction
is neutron as well. This particle can be identified using the missing
mass technique.  We used the energy and scattering angles of the
identified particles to calculate the rest mass of the undetected
particle. Figure~\ref{missmass} shows a typical result of the
calculated mass of the undetected particle using the energy, polar and
azimuthal angles of the detected proton and deuteron. The scattering
angle of both detected particles is fixed to be $25^{\circ} \pm
2^{\circ}$ and the difference between the azimuthal angles of the two
detected particles is $180^{\circ} \pm 5^{\circ}$. The dashed (solid)
curve represents the results of the calculated missing particle mass
before (after) applying the identification cut on
Fig.~\ref{TOF_TDC_E}. After applying the identification cut the mass
spectra show a clear peak corresponding to the mass of neurons with a
FWHM of 6.38~MeV which confirms the accuracy of PID procedure.  The
events on the right-hand side of the peak correspond to break-up
events that undergo a hadronic interaction in the scintillator. These
are discarded in the analysis and properly accounted for in the final
analysis of cross sections with the help of simulations.

The next step in the event selection for the three-body break-up
channel is to find the energy correlation between the final-state
protons and deuterons for a particular kinematical configuration
($\theta_{d}, \theta_{p}, \phi_{12}$).  The number of break-up events
in an interval $S-\frac{\Delta S}{2}$, and $S+\frac{\Delta S}{2}$ was
obtained by projecting the events on a line perpendicular to the
$S$-curve ($D$-axis). The value of $\Delta S$ was $\pm5$ MeV for the
forward wall data.
\begin{figure}[!h]
\centering
\includegraphics[width=1\columnwidth,angle=0]{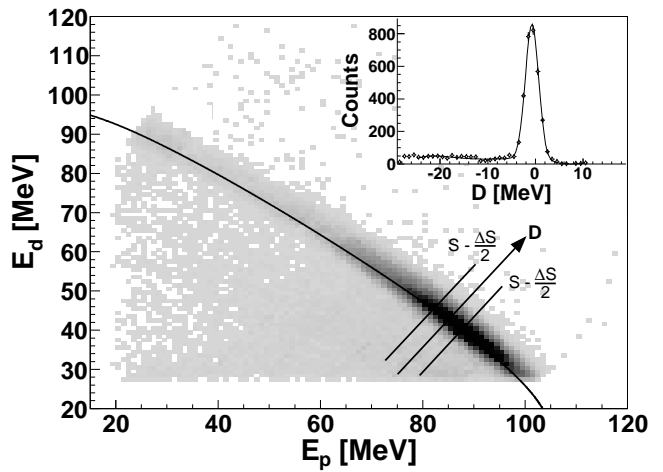}
\caption{The correlation between the energies of the deuteron and the
  proton originating from the three-body break-up channel in one selected
  configuration. The projection of the events from one $S$-bin onto the 
  $D$-axis is shown in the inset.}
\label{25-25-180}       
\end{figure}
Figure~\ref{25-25-180} depicts the correlation between the energy of
protons and deuterons in coincidence for the kinematical
configuration, ($\theta_{d}, \theta_{p}, \phi_{12})=(25^{\circ},
25^{\circ},180^{\circ})$.  The solid curve is the expected correlation
for this configuration. One of the many $S$-intervals and the
corresponding $D$-axis are also shown. The result of the projection of
events on the $D$-axis for a particular $S$-bin is presented in the
inset of Fig.~\ref{25-25-180}. This spectrum consists of mainly
break-up events with a negligible amount of accidental
background. Most of the particles of the break-up events deposit all
their energy in the scintillator, which gives rise to a peak around
zero in the variable $D$. The events on the left-hand side of the peak
at zero correspond to break-up events that undergo a hadronic
interaction in the scintillator (same as those on the right side of
the peak in Fig.~\ref{missmass}).

The interaction of a polarized beam with an unpolarized target
produces an azimuthal asymmetry or an azimuthally uniform change in
the scattering cross section. The magnitude of this effect is
proportional to the product of the polarization of the beam and an
observable that is called the analyzing power. The expression for the
cross section of any reaction induced by a polarized spin-1 projectile
is~{\cite{RamazaniA_Ref32,RamazaniA_Ref33}}:
\begin{eqnarray}
\sigma(\xi, \phi) = \sigma_0(\xi)\lbrack1&+&\sqrt{3}p_Z\rm {Re}(iT_{11}(\xi))\cos\phi \nonumber\\
&-&\frac{1}{\sqrt{8}}p_{ZZ}\rm {Re}(T_{20}(\xi))\nonumber\\
&-&\frac{\sqrt{3}}{2}p_{ZZ}\rm {Re}(T_{22}(\xi))\cos 2\phi \rbrack,
\label{ddbreakcrosFurmola}
\end{eqnarray}
where $\sigma$ and $\sigma_{0}$ are the polarized and unpolarized
cross sections, respectively, and $\xi$ represents the configuration
$(\theta_{d}, \theta_{p}, \phi_{12}, S)$. Note that
Eq.~\ref{ddbreakcrosFurmola} does not contain terms with ${\rm
  Im}(iT_{11})$, ${\rm Re}(T_{21})$, ${\rm Im}(T_{21})$, ${\rm
  Im}(T_{20})$ and ${\rm Im}(T_{22})$. These contributions vanish
because we took explicitly $\beta=90^\circ$ and $\phi_{12}$ is the
absolute value of the difference between the azimuthal angles of the
two outgoing particles. The angle $\beta$ is the angle between the
polarization axis and the momentum of the incoming beam.  In this
work, the variables $\rm {Re}(iT_{11})$, $\rm {Re}(T_{20})$ and $\rm
{Re}(T_{22})$ will be referred to as $iT_{11}$, $T_{20}$ and $T_{22}$,
respectively.  The quantities $iT_{11}$ and $p_{Z}$ are the
vector-analyzing power and the vector beam polarization,
respectively. The observables $T_{20}$ and $T_{22}$ are the
tensor-analyzing powers, $p_{ZZ}$ is the tensor polarization of the
beam, and $\phi$ is the azimuthal scattering angle of the deuteron.
\begin{figure}[!h]
\centering
\includegraphics[width=1.15\columnwidth,angle=0]{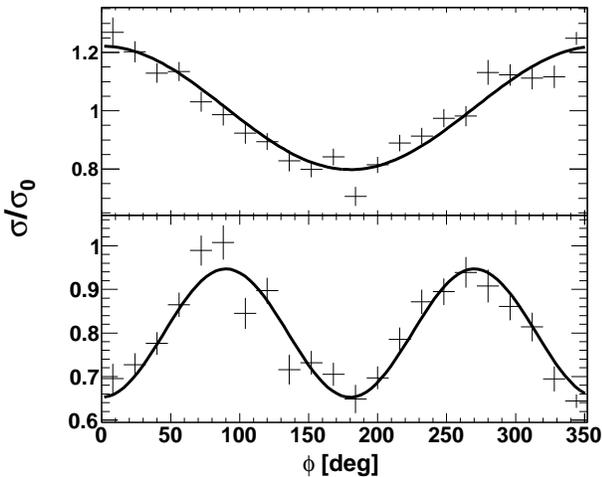}
\caption{The ratio of the spin-dependent cross
section to the unpolarized one for a pure vector-polarized deuteron
beam (top panel) and a pure tensor-polarized deuteron beam (bottom
panel) for ($\theta _{d}=25^\circ , \theta _{p}=25^\circ , \phi _{12}=180^\circ, S=230$~MeV).}
\label{ddBAsy}       
\end{figure}

For a deuteron beam with a pure vector polarization, the ratio $\frac
{\sigma}{\sigma _{0}}$ should show a $\cos \phi$ distribution where
$\sigma$ and $\sigma_{0}$ are the polarized and unpolarized cross
sections, respectively. When a pure tensor-polarized deuteron beam is
used, the ratio $\frac {\sigma}{\sigma _{0}}$ should show a $\cos
2\phi$ distribution. These asymmetries are exploited to extract the
vector-analyzing power, $iT_{11}$ and the tensor-analyzing powers,
$T_{20}$ and $T_{22}$, for every kinematical configuration,
$(\theta_{d},\theta_{p},\phi_{12},S)$.
\begin{figure*}[!th]
\centering
\includegraphics[width=2\columnwidth,angle=0]{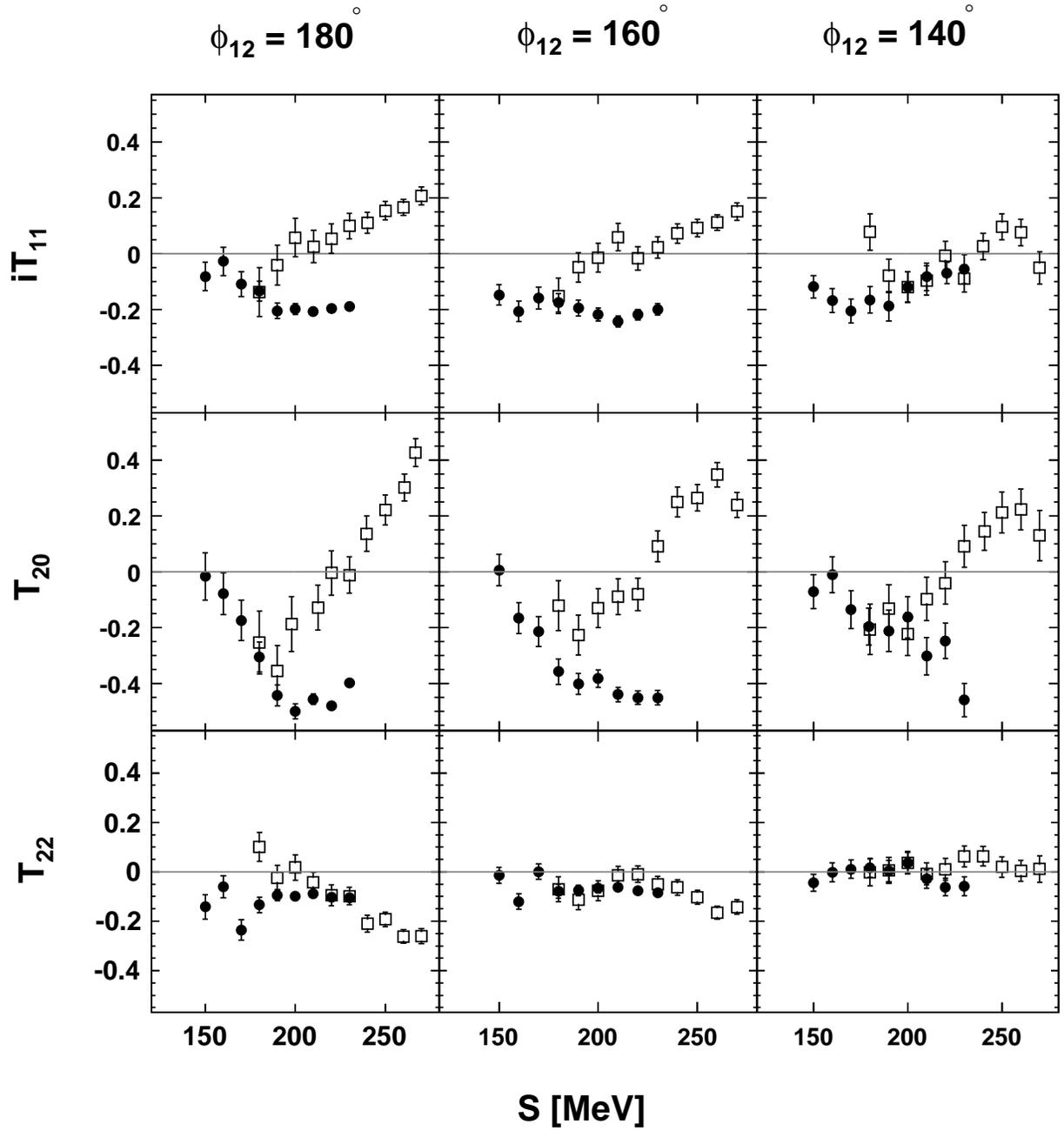}
\caption{The vector- and tensor-analyzing powers at $(\theta_{d},
  \theta_{p}) = (15^{\circ}, 15^{\circ})$ (open squares) and
  $(\theta_{d}, \theta_{p}) = (25^{\circ}, 25^{\circ})$ (filled
  circles) as a function of $S$ for different azimuthal opening
  angles.  The gray lines in all panels show the zero level of the
  analyzing powers.  Only statistical uncertainties are indicated.}
\label{dd1}
 \end{figure*}
\section{Experimental results}
For a measurement of the analyzing powers, we compare the distribution
of the scattered particles at different azimuthal angles for each
polarization state with that obtained with the unpolarized beam. The
$\phi$-distribution obtained with the polarized beam is normalized to
that obtained with an unpolarized beam to obtain the ratio $\frac
{\sigma}{\sigma _{0}}$. This ratio depends on the the number of counts
under the peak after background subtraction and the total integrated
charge collected in the Faraday cup. With this normalization other
parameters like the geometrical asymmetries, inefficiencies, and
target thickness are eliminated.  Figure~\ref{ddBAsy} shows the ratio
$\frac {\sigma}{\sigma _{0}}$ for a pure vector-polarized deuteron
beam (top panel) and a pure tensor-polarized deuteron beam (bottom
panel) for ($\theta_{p}=25^\circ$, $\theta_{d}=25^\circ$,
$\phi_{12}=180^\circ$, $S=230$~MeV).  The curves in the top and bottom
panels are the results of a fit based on Eq.~\ref{ddbreakcrosFurmola}
through the obtained asymmetry distribution for a beam with a pure
vector and tensor polarization, respectively.  The amplitude of the
$\cos\phi $ modulation in the top panel equals to
$\sqrt{3}p_{Z}iT_{11}$ and that of the $\cos 2\phi $ modulation in the
lower panel equals to $- \frac {\sqrt{3}}{2}p_{ZZ}T_{22}$. The offset
from 1 in the lower panel is $- \frac {1}{\sqrt{8}}p_{ZZ}T_{20}$.  The
vector- and tensor-analyzing powers for a few kinematical
configurations of the three-body break-up reaction were extracted.
Figure~\ref{dd1} represents the vector- and tensor-analyzing powers at
$(\theta_{d}, \theta_{p}) = (15^{\circ}, 15^{\circ})$ (open squares)
and $(\theta_{d}, \theta_{p}) = (25^{\circ}, 25^{\circ})$ (filled
circles) as a function of $S$ for different azimuthal opening
angles. The gray lines in all panels show the zero level of the
analyzing powers.  Only statistical uncertainties are indicated.  The
total systematic uncertainty for the analyzing power is estimated to
be $\sim 7.5\%$ which mainly stems from the uncertainty in the
  measurement of the beam polarization via elastic scattering and to a
  much lesser extent from the error of the beam-current correction in
  the analysis of the T20 in the elastic d + d channel. 

This paper demonstrates for the first time the feasibility of
obtaining precision data of the three-body break-up channel in
deuteron-deuteron scattering at the energy of 65~MeV/nucleon.  The
three-body break-up reaction has been clearly identified using the
measured scattering angles, energies, and TOF of the final-state
protons and deuterons. In this work, we analyzed a part of the data in
which the protons and deuterons were scattered into the forward wall
of BINA.  We have provided precision data for the vector and tensor
analyzing powers of the three-body break-up reaction for an incident
deuteron beam of 65~MeV/nucleon~\cite{RamazaniA_Ref30}. These
four-body scattering experiments at KVI will provide in the near
future a complete database in deuteron-deuteron scattering at
intermediate energies including the elastic, transfer, three-body
break-up, and four-body break-up channels.  Together with the upcoming
state-of-the-art ab-initio
calculations~{\cite{Deltu07_1,Deltu07_2,Deltu09}}, these data will
provide the basis to understand the mechanisms behind many-body force
effects.

\section{ Acknowledgments}
The authors acknowledge the work by the cyclotron and ion-source
groups at KVI for delivering a high-quality beam used in these
measurements. This work was performed as part of the research program
of the ``Stichting voor Fundamenteel Onderzoek der Materie''
(FOM). Furthermore, the present work has been performed with financial
support from the University of Groningen (RuG), the Helmholtzzentrum
f\"ur Schwerionenforschung GmbH (GSI), Darmstadt, the Nederlandse
Organisatie voor Wetenschappelijk Onderzoek (NWO), the Polish
2008-2011 science founds as a research project No.~N N202 078135 and
Iran Ministry of Science, Research, and Technology, Kashan
university. CDB and EJS acknowledge support from the US National
Science Foundation through grant PHY 04-57219.


\vspace{15mm}
\begin{thebibliography}{9}
\bibitem{Yukawa}
H.~Yukawa,    \rm  Proc. Phys. Math. Soc. Jap. \textbf{17}, 48 (1935).
\bibitem{RamazaniA_Ref2}
L. D.~Faddeev,  \rm    Sov. Phys. JETP.  \textbf{12}, 1014 (1961).
\bibitem{RamazaniA_Ref3}
W. Gl{\"o}ckle, H.  Wita\l{}a, D. H{\"u}ber, H. Kamada and J. Golak,  \rm    Physics Reports \textbf{274}, 107 (1996).
\bibitem{RamazaniA_Ref4}
R. B.~Wiringa,  A. R. Smith,  and T. L. Ainsworth,   \rm  Phys. Rev. C. \textbf{29}, 1207 (1984).
\bibitem{RamazaniA_Ref5}
H.~Primakoff,  and T. Holstein,   \rm  Phys. Rev.  \textbf{55}, 1218 (1939).
\bibitem{Wit98}
 H.~Wita\l{}a,  W. Gl{\"o}ckle, D. H\"uber, J. Golak and H. Kamada   \rm   Phys. Rev. Lett.  \textbf{81}, 1183 (1998).
\bibitem{RamazaniA_Ref9} H.~Shimizu \textit{et al.}, \rm Nucl. Phys. A
  \textbf{382}, 242 (1982).
\bibitem{RamazaniA_Ref10}
H.~Sakai \textit{et al.},   \rm  Phys. Rev. Lett. \textbf{84}, 5288 (2000).
\bibitem{RamazaniA_Ref11}
K.~Ermisch \textit{et al.},   \rm  Phys. Rev. Lett. \textbf{86}, 5862 (2001).
\bibitem{RamazaniA_Ref12}
K.~Ermisch \textit{et al.},   \rm  Phys. Rev. C \textbf{68}, 051001 (2003).
\bibitem{RamazaniA_Ref13}
H.~ Hatanaka \textit{et al.},   \rm  Eur. Phys. J. A \textbf{18}, 293 (2003).
\bibitem{RamazaniA_Ref14}
E.~Stephan \textit{et al.},   \rm  Phys. Rev. C \textbf{76}, 057001 (2005). 
\bibitem{RamazaniA_Ref15}
K.~Ermisch \textit{et al.},   \rm  Phys. Rev. C \textbf{71}, 064004 (2005).
\bibitem{RamazaniA_Ref16}
K.~Sekiguchi \textit{et al.},   \rm   Phys. Rev. Lett. \textbf{95}, 162301 (2005).
\bibitem{RamazaniA_Ref17}
H.~Mardanpour \textit{et al.},   \rm  Eur. Phys. J. A \textbf{31}, 383 (2007).
\bibitem{Ahmad07}
A.~Ramazani-Moghaddam-Arani \textit{et al.}, \rm Few-body system \textbf{44}, 27 (2008).
\bibitem{Ahmad08}
A.~Ramazani-Moghaddam-Arani \textit{et al.}, \rm Phys. Rev. C \textbf{78}, 014006 (2008).
\bibitem{Wit93}
 H.~Wita\l{}a,  T. Gornelius and W. Gl{\"o}ckle,  \rm    Few-body system \textbf{15}, 67 (1993).
\bibitem{Ed99}
E. J.~Stephenson,  H.  Wita\l{}a, W. Gl{\"o}ckle,  H. Kamada, and A. Nogga  \rm   Phys. Rev. C \textbf{60}, 061001(R) (1999). 
\bibitem{Mesh00}
J. G.~Messchendorp \textit{et al.}, \rm Phys. Lett. B \textbf{481}, 171 (2000).
\bibitem{RamazaniA_Ref18}
R.~Bieber \textit{et al.},   \rm  Phys. Rev. Lett. \textbf{84}, 606 (2000).
\bibitem{Cadman01}
R. V.~Cadman \textit{et al.},   \rm  Phys. Rev. Lett. \textbf{86}, 967 (2001).
\bibitem{Kuros02}
 J.~Kuro\ifmmode \acute{s}\else \'{s}\fi{}-\ifmmode \dot{Z}\else \.{Z}\fi{}o\l{}nierczuk,        \textit{et al.},   \rm  Phys. Rev. C \textbf{66}, 024003 (2002).
\bibitem{Seki02}
K.~Sekiguchi \textit{et al.},   \rm   Phys. Rev. C \textbf{65}, 034003 (2002).
\bibitem{Hatanaka02}
K.~ Hatanaka \textit{et al.},   \rm  Phys. Rev. C \textbf{66}, 044002 (2003).
\bibitem{Meyer04}
H. O.~Meyer \textit{et al.},   \rm  Phys. Rev. Lett. \textbf{93}, 112502 (2004).
\bibitem{RamazaniA_Ref19}
K.~Sekiguchi \textit{et al.},   \rm  Phys. Rev. C \textbf{70}, 014001 (2004).
\bibitem{Mehman05}
 A.A.~Mehmandoost-Khajeh-Dad \textit{et al.}, \rm Phys. Lett. B \textbf{617}, 18 (2005).
\bibitem{Przew06}
B. v.~Przewoski \textit{et al.},   \rm  Phys. Rev. C \textbf{74}, 064003 (2006). 
\bibitem{Kist06}
S.~Kistryn \textit{et al.},   \rm  Phys.  Lett. B \textbf{641}, 23 (2006).
 \bibitem{Bieg06}
A.~Biegun \textit{et al.},   \rm  Acta Phys.  Pol. B \textbf{371}, 213 (2006).
\bibitem{RamazaniA_Ref21}
H.~Amir-Ahmadi  \textit{et al.},   \rm  Phys. Rev. C \textbf{75}, 041001 (2007).
\bibitem{Stephan07}
E.~Stephan \textit{et al.},   \rm  Phys. Rev. C \textbf{76}, 057001 (2007). 
\bibitem{RamazaniA_Ref20}
H.~Mardanpour, Ph.D. thesis,   \rm  University of Groningen (2008).

\bibitem{RamazaniA_Ref22}
M. Eslami-Kalantari, Ph.D. thesis,   \rm  University of Groningen (2009).
\bibitem{Mardan10}
H.~Mardanpour \textit{et al.},  \rm   Phys. Lett. B \textbf{687}, 149 (2010).

\bibitem{Pieper01}
S. C.~Pieper, V. R. Pandharipande, R. B. Wiringa, and J. Carlson, \rm  Phys. Rev. C  \textbf{64}, 014001 (2001).
\bibitem{AV18}
R. B.~Wiringa, V. G. J. Stoks, and R. Schiavilla, \rm Phys. Rev. C  \textbf{51}, 38 (1995).
\bibitem{Pudliner95}
B. S. Pudliner, V. R. Pandharipande, J. Carlson, and R. B.Wiringa, \rm Phys. Rev. Lett \textbf{74}, 4396 (1995).

\bibitem{RamazaniA_Ref28}
V.~Bechtold,  L.~Friedrich, M. S.~Abdel-Wahab, J.~Bialy, M.~Junge, and  F. K.~Schmidt,   \rm  Nucl. Phys. A \textbf{288}, 189 (1977).
\bibitem{RamazaniA_Ref27}
C.~Alderliesten, A.~Djaloeis, J.~Bojowald,  C.~Mayer-B\"oricke, G.~Paic,  and T.~Sawada,   \rm  Phys. Rev. C \textbf{18}, 2001 (1978).

\bibitem{RamazaniA_Ref29}
M. Gar\c{c}on \textit{et al.},   \rm  Nucl. Phys. A \textbf{458}, 287 (1986).

\bibitem{Mich2007}
A. M. Micherdzi\ifmmode \acute{s}\else \'{n}ska, \textit{et al.},  Phys. Rev. C \textbf{75}, 054001 (2007).

\bibitem{RamazaniA_Ref23}
F.~Ciesielski and J.~Carbonell and C.~Gignoux,   \rm   Phys. Lett. B \textbf{447}, 199 (1999).
\bibitem{RamazaniA_Ref24}
A. C.~Fonseca,   \rm   Phys. Rev. Lett \textbf{83}, 4021 (1999).
\bibitem{RamazaniA_Ref25}
M.~Viviani, A.~Kievsky, S.~Rosati, E. A.~George,  and L. D.~Knutson,   \rm  Phys. Rev. Lett \textbf{86}, 3739 (2001).
\bibitem{RamazaniA_Ref26}
R.~Lazauskas,  J.~Carbonell, A. C.~Fonseca, M.~Viviani, A.~Kievsky, and S.~Rosati,   \rm  Phys. Rev. C \textbf{71}, 034004 (2005).

\bibitem{RamazaniA_Ref30}
A. Ramazani-Moghaddam-Arani, Ph.D. thesis,   \rm   University of Groningen (2009).


\bibitem{Polis}
H. R. Kremers and A. G. Drentje, Vol. 421 of AIP Conf. Proc., edited by R. J. Holt and M. A. Miller (AIP, New York,1997), p. 507.
\bibitem{RamazaniA_Ref31}



N. Kalantar-Nayestanaki, J. Mulder and J. Zijlstra,   \rm  Nucl. Instr. and Meth. in Phys. Res. A \textbf{417}, 215 (1998).
\bibitem{Kremers1}
H. R.~Kremers, J.P.M.~Beijers, N.~Kalantar-Nayestanaki, and T.B.~Clegg,  \rm  Nucl. Instr. and Meth. in Phys. Res. A \textbf{516}, 209 (2004).

\bibitem{BBS_ref}
A. M. van den Berg,  \rm  Nucl. Instr. and Meth. in Phys. Res. B \textbf{99}, 637 (1995).

\bibitem{RamazaniA_Ref32}
G. G. Ohlsen,   \rm  Rep. Prog. Phys. \textbf{35}, 717 (1972).
\bibitem{RamazaniA_Ref33}
G. G. Ohlsen,   \rm  Nucl. Instr. Meth. \textbf{179}, 283 (1981).
\bibitem{Deltu07_1}
A. Deltuva,   and A. C. Fonseca,  Phys. Rev. C \textbf{76},  021001(R) (2007).
\bibitem{Deltu07_2}
A. Deltuva,   and A. C. Fonseca,  Phys. Rev. C \textbf{75},   014005 (2007).
\bibitem{Deltu09}
A. Deltuva, Private communication.


 \end{thebibliography}
\end{document}